\documentclass[aps,english,prl,amsmath,amsfonts,amssymb,superscriptaddress,twocolumn,showpacs,citeautoscript]{revtex4-1}
\usepackage[T1]{fontenc}
\usepackage[latin9]{inputenc}
\usepackage{amsmath}
\usepackage{amssymb}
\usepackage{wasysym}
\usepackage{esint}
\usepackage{graphicx}
\usepackage{times}
\usepackage{nicefrac}
\usepackage{appendix}
\usepackage{textcase}
\usepackage{subfigure}
\usepackage{empheq}
\usepackage{color}
\usepackage{leftidx}
\usepackage{makecell}
\usepackage{stackrel}
\makeatletter
\@ifundefined{textcolor}{}
{%
 \definecolor{BLACK}{gray}{0}
 \definecolor{WHITE}{gray}{1}
 \definecolor{RED}{rgb}{1,0,0}
 \definecolor{GREEN}{rgb}{0,1,0}
 \definecolor{BLUE}{rgb}{0,0,1}
 \definecolor{CYAN}{cmyk}{1,0,0,0}
 \definecolor{MAGENTA}{cmyk}{0,1,0,0}
 \definecolor{YELLOW}{cmyk}{0,0,1,0}
}

\renewcommand{\vec}[1]{\mathbf{#1}}

\renewcommand{\Im}{\operatorname{Im}}

\renewcommand{\o}[1]{{#1}}

\renewcommand{\b}{\beta}

\newcommand{\add}[1]{\if\a\b{{\color{red} #1}}\else{#1}\fi}

\newcommand{\bracket}[1]{\langle #1 \rangle}
\newcommand{\ket}[1]{| #1 \rangle}
\newcommand{\bra}[1]{\langle #1 |}
\newcommand{\im}{\mathrm{i}}

\newcommand{\citeasnoun}[1]{Ref.~\onlinecite{#1}}

\renewcommand{\eqref}[1]{(\ref{eq:#1})}

\newcommand{\figref}[1]{Fig.~\ref{fig:#1}}
\newcommand{\Figref}[1]{Figure~\ref{fig:#1}}

\newcommand{\Tr}{\text{Tr }}

\makeatother
\usepackage{babel}

\begin{document}
\title{Unifying microscopic and continuum treatments of van der Waals and Casimir interactions}

\author{Prashanth S. Venkataram}
\affiliation{Department of Electrical Engineering, Princeton University, Princeton, New Jersey 08544, USA}
\author{Jan Hermann}
\affiliation{Fritz-Haber-Institut der Max-Planck-Gesellschaft, Faradayweg 4--6, 14195, Berlin, Germany}
\author{Alexandre Tkatchenko}
\affiliation{Fritz-Haber-Institut der Max-Planck-Gesellschaft, Faradayweg 4--6, 14195, Berlin, Germany}
\affiliation{Physics and Materials Science Research Unit, University of Luxembourg, L-1511 Luxembourg}
\author{Alejandro W. Rodriguez}
\affiliation{Department of Electrical Engineering, Princeton University, Princeton, New Jersey 08544, USA}

\date{\today}

\begin{abstract}
  We present an approach for computing long-range van der Waals (vdW)
  interactions between complex molecular systems and arbitrarily
  shaped macroscopic bodies, melding atomistic treatments of
  electronic fluctuations based on density functional theory in the
  former, with continuum descriptions of strongly shape-dependent
  electromagnetic fields in the latter, thus capturing many-body and
  multiple scattering effects to all orders. Such a theory is
  especially important when considering vdW interactions at mesoscopic
  scales, i.e. between molecules and structured surfaces with features
  on the scale of molecular sizes, in which case the finite sizes,
  complex shapes, and resulting nonlocal electronic excitations of
  molecules are strongly influenced by electromagnetic retardation and
  wave effects that depend crucially on the shapes of surrounding
  macroscopic bodies. We show that these effects together can modify
  vdW interactions by orders of magnitude compared to previous
  treatments based on Casimir--Polder or non-retarded approximations,
  which are valid only at macroscopically large or atomic-scale
  separations, respectively.
\end{abstract}

\maketitle

% Dispersive van der Waals (vdW) interactions are pervasive in molecular systems, especially those of interest to biologists, biophysicists, and other soft-matter scientists
Van der Waals (vdW) interactions play an essential role in
non-covalent phenomena throughout biology, chemistry, and
condensed-matter physics~\cite{WoodsRMP2015,
  Langbein1974,TkatchenkoADFM2014}. It has long been known that vdW
interactions among a system of polarizable atoms are not
pairwise-additive but instead strongly depend on geometric and
material properties~\cite{McLachlanMP1963, Langbein1974,
  ColeMS2009}. However, only recently developed theoretical methods
have made it possible to account for short-range quantum interactions
in addition to long-range many-body screening in molecular
ensembles~\cite{TkatchenkoJCP2013, GobreNCOMMS2013, DiStasioJPCM2014,
  TkatchenkoADFM2014, AmbrosettiSCIENCE2016, PhanJAP2013,
  ReillyCS2015, ShtogunJPCL2010, AmbrosettiJCP2014, KimLANGMUIR2007,
  TkatchenkoPRL2012}, demonstrating that nonlocal many-body effects
cannot be captured by simple, pairwise-additive descriptions; these
calculations typically neglect electromagnetic retardation effects in
molecular systems. Simultaneously, recent theoretical and experimental
work has characterized dipolar Casimir--Polder interactions between
macroscopic metallic or dielectric objects and atoms, molecules, or
Bose--Einstein condensates, further extending to nonzero temperatures,
dynamical situations, and fluctuations in excited states (as in
so-called Rydberg atoms)~\cite{BuhmannPRA2012, BuhmannIJMPA2016,
  BenderPRX2014, ThiyamPRE2014, BarcellonaPRA2016, Intravaia2011,
  DeKieviet2011, BabbJPCS2005, Buhmann2012I, Buhmann2012II}. Yet,
while theoretical treatments have thus far accounted for the full
electrodynamic response of macroscopic bodies (including retardation),
they often treat molecules as point dipoles of some effective bulk
permittivities or as collections of noninteracting atomic dipoles,
ignoring finite size and other many-body electromagnetic effects.

In this paper, motivated by the aforementioned theoretical
developments~\cite{Johnson2011, BuhmannPRA2012, BuhmannIJMPA2016,
  BenderPRX2014, Buhmann2012I, Buhmann2012II, WoodsRMP2015,
  RodriguezADP2015, RodriguezNATURE2011}, we describe an approach that
seamlessly connects atomistic descriptions of large molecules to
continuum descriptions of \emph{arbitrary} macroscopic bodies,
characterizing their mutual vdW interactions. In particular, while
molecules that are very close to macroscopic objects require atomistic
descriptions of the latter, and very large molecules that are far from
macroscopic objects require consideration of the contributions of
vibrational (in addition to electronic) resonances to the vdW
interaction energy, we focus on a mesoscopic regime involving
molecular sizes and separations on the order of 1--100 $\mathrm{nm}$,
where macroscopic objects can be treated continuously for the purposes
of computing electromagnetic field responses (and molecular
vibrational resonances can be neglected), yet electromagnetic
retardation in conjunction with the finite sizes, nontrivial shapes,
and nonlocal electronic correlations of large molecules need to be
self-consistently considered to accurately characterize vdW
interactions. We specifically investigate interactions among various
large molecules and gold surfaces, and show that the effect of
nonlocal polarization correlations, encapsulated in the ratio of
retarded, many-body (RMB) to pairwise vdW energies (or forces), causes
relative deviations from pairwise treatments ranging from 20\% to over
3 orders of magnitude; further differences of over an order of
magnitude are observed when retardation or finite size effects are
neglected. % Our method is general enough to handle molecules and
%macroscopic bodies of arbitrary geometries and materials.

\begin{figure}[t!]
\centering
\includegraphics[width=0.9\columnwidth]{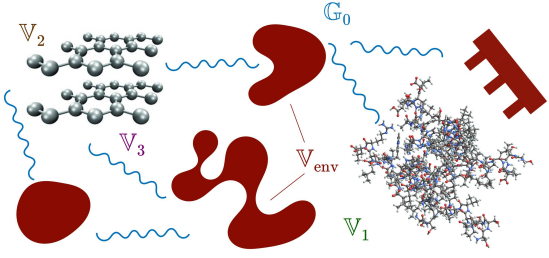}
\caption{Schematic of molecular bodies described by electric
  susceptibilities $\mathbb{V}_{n}$ in the vicinity of and interacting
  with macroscopic bodies described by a collective susceptibility
  $\mathbb{V}_{\mathrm{env}}$, where the interactions are mediated by
  vacuum electromagnetic fields $\mathbb{G}_{0}$.}
\label{fig:schematic}
\end{figure}

The basis of our work is an equation for the long-range dispersive vdW
energy of a system of polarizable bodies, consisting of $N$
microscopic bodies (molecules), labeled by $k$ and described by
electric susceptibilities $\mathbb{V}_{k}$, and a collection of
continuum bodies (an environment) described by a collective,
macroscopic susceptibility $\mathbb{V}_{\mathrm{env}}$, displayed
schematically in \figref{schematic}. The energy of such a collection
of bodies can be obtained from the scattering
framework~\cite{RahiPRD2009} and written as an integral over imaginary
frequency $\omega = \im\xi$,
\begin{equation} \label{eq:TT}
  \mathcal{E} = \frac{\hbar}{2\pi}
  \int_{0}^{\infty} \mathrm{d}\xi~\ln[\det(\mathbb{T}_{\infty}
  \mathbb{T}^{-1})],
\end{equation}
in terms of T-operators that depend on the bodies' susceptibilities as
well as on the homogeneous electric Green's function
$\mathbb{G}_{0}(\im\xi, \vec{x}, \vec{x}') = (\nabla \otimes \nabla -
\frac{\xi^{2}}{c^{2}} \mathbb{I})\frac{e^{-\frac{\xi |\vec{x} -
      \vec{x}'|}{c}}}{4\pi |\vec{x} - \vec{x}'|}$ (including
retardation) mediating electromagnetic interactions; they encode the
scattering properties of the various bodies, and are given by,
\begin{equation*}
  \mathbb{T} = \left(\mathbb{I} - (\mathbb{V} + \mathbb{V}_{\mathrm{env}}) \mathbb{G}_{0}\right)^{-1} (\mathbb{V} + \mathbb{V}_{\mathrm{env}}),
\end{equation*}
where $\mathbb{V} = \sum_k \mathbb{V}_k$; $\mathbb{T}_{\infty} =
\mathbb{T}_{\mathrm{env}} \prod_{k} \mathbb{T}_{k}$, written in terms
of $\mathbb{T}_{k(\mathrm{env})} = (\mathbb{I} -
\mathbb{V}_{k(\mathrm{env})} \mathbb{G}_{0})^{-1}
\mathbb{V}_{k(\mathrm{env})}$, encodes the scattering response of the
bodies in isolation from one another~\cite{RahiPRD2009}.

The energy in~\eqref{TT} treats microscopic and macroscopic bodies on
an equal footing, yet the key to its accurate evaluation lies in
appropriately representing the degrees of freedom (DOFs) of each
entity. Typically, macroscopic environments are well described by
continuum susceptibilities $\mathbb{V}_{\mathrm{env}}$, whose response
can be expanded in a basis of incoming and outgoing propagating
planewaves, as is typical of the scattering
framework~\cite{RahiPRD2009}, or via localized functions,
e.g. tetrahedral mesh elements, in brute-force
formulations~\cite{ReidPRA2013,RodriguezADP2015}. Microscopic bodies,
on the other hand, generally require quantum descriptions, but recent
work has shown that one can accurately represent their response
$\mathbb{V}_{k} = \sum_p \alpha_p \ket{f_p} \bra{f_p}$ through bases
$\{\ket{f_{p}}\}$ of either exponentially localized (for insulators)
or polynomially delocalized (for metals) functions~\cite{GePRB2015},
that accurately capture multipolar interactions among electronic
wavefunctions~\cite{TkatchenkoJCP2013, DiStasioJPCM2014,
  TkatchenkoADFM2014, AmbrosettiJCP2014}. For molecules with finite
electronic
gaps, %the electron densities are well described by sums over dipolar ground-state oscillator wavefunctions comprising bare atomic response functions
the bare response is well described by sums over dipolar ground-state
oscillator densities~\cite{DonchevJCP2006, AmbrosettiSCIENCE2016,
  DiStasioJPCM2014, PhanJAP2013, ShtogunJPCL2010, KimLANGMUIR2007,
  ColeMS2009},
\begin{equation}
  f_{p} (\mathrm{i}\xi, \vec{x}) = \left(\sqrt{2\pi}
    \sigma_{p} \left(\mathrm{i}\xi\right)\right)^{-3}
  \mathrm{exp}\left(-\frac{ (\vec{x} - \vec{x}_{p})^2 }{2\
      \sigma_{p}^{2} \left(\mathrm{i}\xi\right)}\right),
\end{equation}
centered at the locations $\vec{x}_{p}$ of each atom $p$, normalized
such that $\int \mathrm{d}^{3} \vec{x}~f_{p} = 1$, and featuring a
Gaussian width that, rather than being
phenomenological~\cite{MahantyJCSFT1975, RennePHYSICA1971A}, depends
on the atomic polarizability via $\sigma_{p} (\mathrm{i}\xi) =
\left(\frac{\alpha_{p}
    (\mathrm{i}\xi)}{\sqrt{72\pi^3}}\right)^{1/3}$~\cite{MayerPRB2007,
  DiStasioJPCM2014}. The isotropic atomic polarizabilities
$\alpha_{p}$ are computed via density functional theory, as in recent
works~\cite{AmbrosettiSCIENCE2016, DiStasioJPCM2014}, which include
short-range electrostatic, hybridization, and quantum exchange
effects.

Since microscopic and macroscopic bodies are assumed to be disjoint,
it is more efficient to partition the T-operators into blocks
belonging to either molecules or macroscopic objects, allowing a trace
over the macroscopic DOFs. The definitions of
$\mathbb{T}_{k(\mathrm{env})}$ imply
$\mathbb{T}_{k(\mathrm{env})}^{-1} = \mathbb{V}_{k(\mathrm{env})}^{-1}
- \mathbb{G}_{0}$, which means that the relevant T-operators can be
written as:
\begin{align}
\mathbb{T}^{-1} &= \begin{bmatrix}
  \mathbb{T}_{\mathrm{mol}}^{-1} & -\mathbb{G}_{0} \\
  -\mathbb{G}_{0} & \mathbb{T}_{\mathrm{env}}^{-1}
\end{bmatrix}, \,\,\,
  \mathbb{T}_{\infty} = \begin{bmatrix}
    \mathbb{T}_{\mathrm{mol},\infty} & 0 \\
    0 & \mathbb{T}_{\mathrm{env}}
\end{bmatrix}
\end{align}
thus partitioning the molecular and macroscopic (environmental)
DOFs. These depend on the molecular T-operators
\begin{align}
  \mathbb{T}_{\mathrm{mol}}^{-1} &= \begin{bmatrix}
    \mathbb{T}_{1}^{-1} & -\mathbb{G}_{0} & \ldots & -\mathbb{G}_{0} \\
    -\mathbb{G}_{0} & \mathbb{T}_{2}^{-1} & \ldots & -\mathbb{G}_{0} \\
    \vdots & \vdots & \ddots & \vdots \\
    -\mathbb{G}_{0} & -\mathbb{G}_{0} & \ldots &\mathbb{T}_{N}^{-1},
  \end{bmatrix}
\end{align}
with $\mathbb{T}_{\mathrm{mol},\infty} = \prod_k \mathbb{T}_k$, which
are in turn partitioned into blocks for each of the $N$ molecular
bodies. Given this, the product in the determinant can be evaluated
as:
\begin{align}
  \det(\mathbb{T}_{\infty} \mathbb{T}^{-1}) &=
  \det(\mathbb{T}_{\mathrm{mol},\infty}
  \mathbb{T}_{\mathrm{mol}}^{-1}) \det(\mathbb{I} - \mathbb{G}_{0}
  \mathbb{T}_{\mathrm{env}}
  \mathbb{G}_{0} \mathbb{T}_{\mathrm{mol}}) \nonumber \\
  &= \det(\mathbb{T}_{\mathrm{mol},\infty}
  \mathbb{T}_{\mathrm{mol}}^{-1}) \det(\mathbb{I} -
  \mathbb{G}_{\mathrm{env}} \mathbb{V}) \nonumber
  \\&\hspace{1.03in}\times\det (\mathbb{I} - \mathbb{G}_{0}
  \mathbb{V})^{-1}
\end{align}
where we used the property $\mathbb{G}_{0}
\mathbb{T}_{k(\mathrm{env})} = (\mathbb{I} - \mathbb{G}_{0}
\mathbb{V}_{k(\mathrm{env})})^{-1} - \mathbb{I}$, and consolidated the
scattering properties of the macroscopic bodies into the operator
$\mathbb{G}_{\mathrm{env}} = \mathbb{G}_{0} (\mathbb{I} -
\mathbb{V}_{\mathrm{env}} \mathbb{G}_{0})^{-1}$, which solves
\begin{equation}
  \left[\nabla \times \nabla \times +
    \frac{\xi^{2}}{c^{2}} \left(\mathbb{I} + \mathbb{V}_{\mathrm{env}} \right)\right]
  \mathbb{G}_{\mathrm{env}} = -\frac{\xi^{2}}{c^{2}} \mathbb{I} 
\end{equation}
for an imaginary frequency $\omega = \im\xi$, thereby encoding the
macroscopic DOFs purely in the electric field response; this can be
solved via any number of state-of-the-art analytical or numerical
classical electrodynamic techniques~\cite{Johnson2011, WoodsRMP2015,
  RodriguezADP2015, RodriguezNATURE2011}, including but not limited to
scattering~\cite{LambrechtNJP2006, RahiPRD2009, ReidPRA2013} and
finite-difference~\cite{RodriguezPRA2007, RodriguezPRA2009,
  McCauleyPRA2010} methods. Moreover, as the molecules are all
disjoint, then $\det(\mathbb{T}_{\mathrm{mol},\infty}
\mathbb{T}_{\mathrm{mol}}^{-1}) = \det(\mathbb{I} - \mathbb{G}_{0}
\mathbb{V}) \prod_{k} \det (\mathbb{I} - \mathbb{G}_{0}
\mathbb{V}_{k})^{-1}$. Putting all of these identities together yields
the following expression for the energy:
\begin{equation} \label{eq:MM}
  \mathcal{E} = \frac{\hbar}{2\pi}
  \int_{0}^{\infty}\mathrm{d}\xi~\ln[\det\left(\mathbb{M}
    \mathbb{M}_{\infty}^{-1}\right)]
\end{equation}
where $\mathbb{M} = \mathbb{I} - \mathbb{G}_{\mathrm{env}} \mathbb{V}$
and $\mathbb{M}_{\infty} = \prod_{k} (\mathbb{I} - \mathbb{G}_{0}
\mathbb{V}_{k})$.

The above log-determinant formula for the energy includes retardation
by construction and accounts for many-body screening and multiple
scattering to all orders, thereby ensuring full consideration of
finite size, complex shape effects, and collective polarization
excitations (see supplement for an alternate equivalent derivation
including all of these effects). Moreover, existing sophisticated
techniques for modeling molecular and electromagnetic-field responses
come together in the operator products $\mathbb{G} \mathbb{V}_{k}$;
when represented in the $p$-dimensional molecular basis
$\{\ket{f_{p}}\}$, their block matrix elements are of the form:
\begin{equation}
\label{eq:Gelem}
  \bracket{f_{p} | \mathbb{G} \mathbb{V}_{k}  f_{q}} =
  \alpha_{q} \int \mathrm{d}^{3} \vec{x}~\mathrm{d}^{3} \vec{x}'~f_{p}
  (\vec{x}) \mathbb{G} (\vec{x}, \vec{x}') f_{q} (\vec{x}')
\end{equation}
(see supplement for more details). The equivalence of~\eqref{TT}
and~\eqref{MM} captures the seamless unification of ideas and methods
previously confined to either atomistic vdW or continuum Casimir
physics~\cite{FrenchRMP2010}: \eqref{MM} is similar to prior
log-determinant expressions used to describe molecular interactions in
vacuum~\cite{DiStasioJPCM2014}, except that $\mathbb{G}_{0}$ and
$\mathbb{G}_{\mathrm{env}}$ are replaced by nonretarded (quasistatic)
vacuum fields $\mathbb{G}_{0}(\im\xi = 0)$.

We demonstrate the importance of all of these effects by comparing the
vdW energies (or forces) obtained from~\eqref{MM} to those from
pairwise or other approximate treatments in a number of
configurations, consisting of one or two molecules above either a gold
half-space or a conical gold tip. While the Green's function of the
half-plate can be computed analytically~\cite{Novotny2006}, the latter
is computed using brute-force numerical techniques~\cite{Johnson2011,
  WoodsRMP2015, RodriguezADP2015, RodriguezNATURE2011}, with the
dielectric function of gold taken
from~\cite{BuhmannPRA2012}. % Note that while the susceptibility of the continuum gold surface should be spatially dispersive, this spatial dispersion, tied to the need for an atomistic description of the gold surface, has a negligible effect on $\mathbb{G}_{\mathrm{env}}$ for molecule--surface separations of more than a few angstroms. % (confusing, not useful)
We specifically study a $\mathrm{C}_{500}$-fullerene of radius
$1~\mathrm{nm}$, a 250 atom $30~\mathrm{nm}$-long linear carbyne wire,
and a 1944 atom-large
$2.6~\mathrm{nm}~\times~2.9~\mathrm{nm}~\times~5.5~\mathrm{nm}$
protein associated with human Huntington's
disease~\cite{FergusonPNAS2006, RCSBPDB, WWPDB}.

We further compare the RMB energy from~\eqref{MM} to typical
approximations used in the literature: the non-retarded vdW energy
$\mathcal{E}_0$, obtained by evaluating~\eqref{MM} with $\mathbb{G}_0$
and $\mathbb{G}_\mathrm{env}$ replaced by their respective quasistatic
($\im\xi = 0$) responses, and the Casimir--Polder (CP) energy,
\begin{equation}
  \mathcal{E}_\mathrm{CP} = -\frac{\hbar}{2\pi}
  \int_{0}^{\infty}~\mathrm{d}\xi~\Tr \left[\alpha \cdot
    \mathbb{G}_\mathrm{env} \cdot \left(\mathbb{I} + \frac{1}{2} \alpha \cdot
      \mathbb{G}_\mathrm{env}\right) \right]
\label{eq:CP}
\end{equation}
which ignores finite size effects by instead contracting the dressed
susceptibility of the molecular ensemble into effective dipolar
polarizabilities,
\begin{equation*}
  \alpha = \bigoplus_k \sum_{p,q} \langle f_{p} |
  (\mathbb{I} - \mathbb{V}_{k} \mathbb{G}_0)^{-1} \mathbb{V}_{k} f_{q}
  \rangle,
\end{equation*}
thus neglecting higher-order many-body interactions among the
different molecules and surfaces. Finally, we define a pairwise
interaction energy,
\begin{equation}
  \mathcal{E}_{\mathrm{PWS}} = -\frac{\hbar}{2\pi} \int_{0}^{\infty}
  \mathrm{d}\xi~\Tr\left[\sum_{k} \mathbb{V}_{k} \mathbb{G}_{\mathrm{env}}
    \left(\mathbb{I} + \frac{1}{2} \sum_{l \neq k} \mathbb{V}_{l}
      \mathbb{G}_{\mathrm{env}} \right) \right]
\label{eq:PWS}
\end{equation}
which, as in~\eqref{CP}, is obtained as a lowest-order expansion of
\eqref{MM} in the scattering; this captures both finite size and
retardation but ignores all high-order many-body interactions, with
the sums over $k, l$ running over either individual or pairs of
molecules. When comparing non-retarded and CP energies to their
corresponding pairwise approximations, it suffices to take the
quasistatic limit in \eqref{PWS} and to let $(\mathbb{I} -
\mathbb{V}_k \mathbb{G}_0)^{-1} \to \mathbb{I}$ for the effective
polarizability $\alpha$ in \eqref{CP}, respectively.

\begin{figure}[t!]
\centering
\includegraphics[width=0.9\columnwidth]{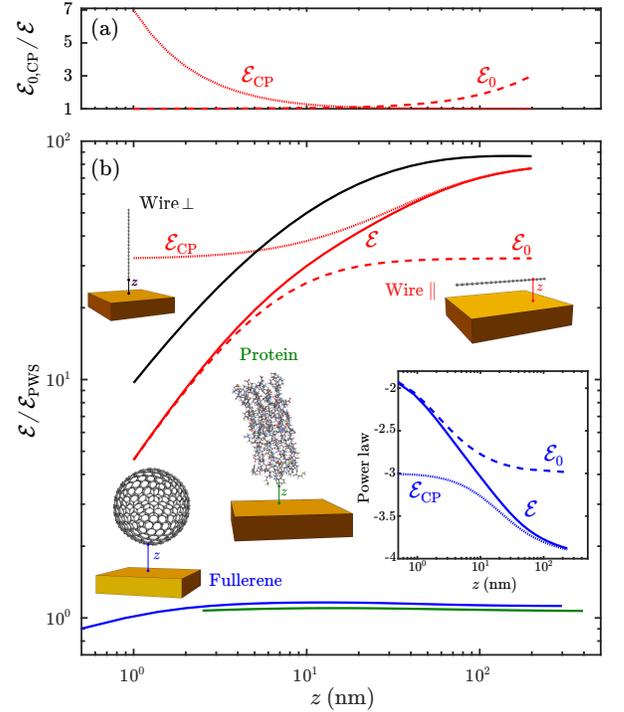}
\caption{(a) CP $\mathcal{E}_{\mathrm{CP}}$ (dotted red) and
  non-retarded $\mathcal{E}_{0}$ (dashed red) energies of a parallel
  carbyne wire separated from a gold plate by a vertical distance $z$,
  normalized to the corresponding retarded, many-body energy
  $\mathcal{E}$ of \eqref{MM}, as a function of $z$; (b) Energy ratio
  $\frac{\mathcal{E}}{\mathcal{E}_{\mathrm{PWS}}}$ versus $z$ for a
  range of molecules, i.e. a fullerene (solid blue), protein (solid
  green), or wire in the parallel (solid red) or perpendicular (solid
  black) orientations, above the gold plate;
  $\mathcal{E}_{\mathrm{PWS}}$ is the energy obtained by a pairwise
  approximation defined in \eqref{PWS}. Also shown are the predictions
  of both CP (dotted red) and non-retarded (dashed red) approximations
  for the case of a parallel wire. Inset: the power law
  $\frac{\partial \ln(\mathcal{E})}{\partial \ln(z)}$ (solid blue) of
  the fullerene--plate system with respect to $z$, compared to both CP
  and non-retarded approximations.}
\label{fig:onemol}
\end{figure}

\Figref{onemol} shows the RMB to pairwise energy ratio
$\frac{\mathcal{E}}{\mathcal{E}_{\mathrm{PWS}}}$ of various
configurations (insets), with the fullerene interaction (blue line)
found to vary only slightly, attaining a maximum of 1.16 at $z \approx
10~\mathrm{nm}$; such a small discrepancy stems from the small size
and isotropic shape of the fullerene, which limits possible nonlocal
correlations in its polarization response. Even weaker relative
correlations are observed in the case of the protein (green line),
which despite its greater size, number of atoms, and chemical
complexity, has a reduced response compared to semi-metallic carbon
allotropes~\cite{AmbrosettiSCIENCE2016, DiStasioJPCM2014}. To separate
the various many-body effects, the inset of \figref{onemol} compares
the RMB power law $\frac{\partial \ln(\mathcal{E})}{\partial \ln(z)}$
of the fullerene interaction to its counterparts when neglecting
either finite size or retardation. As expected, both approximations
become accurate in their corresponding regimes of validity, with the
power law asymptoting to $-4$ and $-1.9$ at large and small $z$,
respectively, but fail in the intermediate, mesoscopic regime $z
\approx 10~\mathrm{nm}$. Even larger discrepancies arise in the case
of the wire, whose large size and highly anisotropic shape support
long-wavelength collective fluctuations. We find that the absolute
values of both $\mathcal{E}_{0}$ (dashed red) and
$\mathcal{E}_{\mathrm{CP}}$ (dotted red) for the parallel wire
overestimate $\mathcal{E}$ by factors of 3--7 [\figref{onemol}(a)] due
to the slower decay of the Green's function in the former and lack of
screening over the length (or modes) of the wire in the latter. The
corresponding energy ratios, however, behave differently in that the
effect of screening is strongest in the quasistatic limit, which ends
up greatly dampening the many-body excitations relative to pairwise
approximations and hence leads to smaller non-retarded energy ratios;
in contrast, by construction CP ignores many-body interactions with
the surface and thus screening has a much weaker impact relative to
the pairwise approximation, leading to larger CP energy ratios. At
intermediate $z \approx 10~\mathrm{nm}$ of the order of the wire
length, $\mathcal{E}/\mathcal{E}_\mathrm{PWS} \approx 30$, with the
approximate energy ratios deviating by 20\%. Similar results are
observed in the case of a wire in the perpendicular orientation (black
lines), with the pairwise energy leading to slightly larger
discrepancies at short separations due to the screening and decreasing
impact of atoms farther away from the plate.

\begin{figure}[t!]
\centering
\includegraphics[width=0.9\columnwidth]{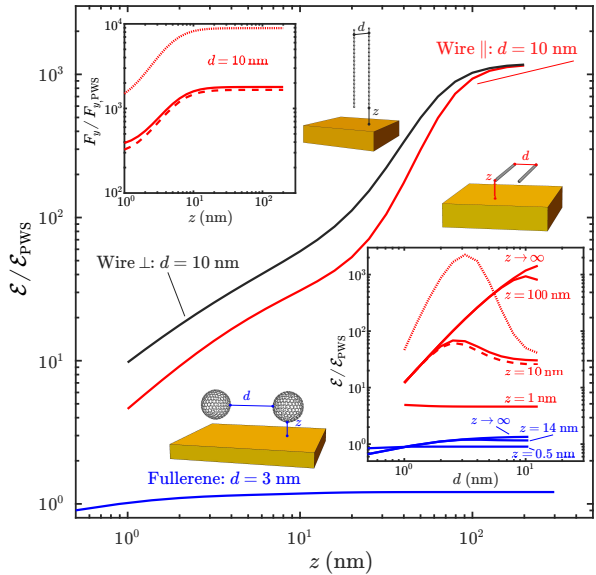}
\caption{Energy ratio $\frac{\mathcal{E}}{\mathcal{E}_{\mathrm{PWS}}}$
  versus vertical distance $z$ for two fullerenes at fixed horizontal
  separation $d = 3~\mathrm{nm}$ (solid blue) or two wires at $d =
  10~\mathrm{nm}$, in either the parallel (solid red) or perpendicular
  (solid black) orientations, above a gold plate. Top inset:
  horizontal-force ratio $\frac{F_{y}}{F_{y,\mathrm{PWS}}}$ versus $z$
  for the parallel wires at $d = 10~\mathrm{nm}$. Bottom inset:
  $\frac{\mathcal{E}}{\mathcal{E}_{\mathrm{PWS}}}$ versus $d$ for the
  fullerenes and the parallel wires at several values of $z$; also
  shown are the corresponding ratios obtained via CP (dotted red) and
  non-retarded (dashed red) approximations, for the particular case of
  $z=10~\mathrm{nm}$.}
\label{fig:twomol}
\end{figure}

We now investigate the mutual vdW interactions among two fullerenes or
parallel wires oriented either parallel or perpendicular to the gold
plate~[\figref{twomol}], focusing primarily on horizontal separations
$d$ on the order of molecular sizes, where many-body and finite size
effects are strongest. Especially in the case of two wires, the
pairwise approximation is shown to fail by many orders of magnitude,
with the largest energy ratios occurring at asymptotically large $z$,
i.e. for two molecules in vacuum, while at small $z$ a decreasing
ratio reflects the dominant interactions (and screening) of the
individual molecules with the plate. The transition and competition
between the two limiting behaviors occurs at mesoscopic $z \sim d$,
and is more clearly visible from the plots in \figref{twomol}(lower
inset), which show $\frac{\mathcal{E}}{\mathcal{E}_{\mathrm{PWS}}}$
versus $d$ at several values of $z$. In particular, in the case of
parallel wires at mesoscopic $z = 10~\mathrm{nm}$, the competition
leads to a nonmonotonic energy ratio, with the maximum of 70 occurring
at intermediate $d \approx 3~\mathrm{nm}$. Comparisons against
non-retarded and CP approximations illustrate behaviors similar to the
previous case of a single wire, with each under- and over-estimating
the ratios by approximately 20\% and 30\%, respectively. Also shown in
\figref{twomol}(upper inset) is the ratio of the horizontal force $F_y
= -\frac{\partial \mathcal{E}}{\partial y}$ on the wires to its
pairwise counterpart, plotted against $z$ for parallel wires at $d =
10~\mathrm{nm}$. Note that by construction, $F_{y,\mathrm{PWS}}$ is
independent of $z$ and thus, the system experiences an absolute
decrease in the force due to the screening induced by the
plate. Comparing $F_{y,0}$ and $F_{y,\mathrm{CP}}$, one finds the
surprising result that in contrast to the energy ratio of a single
molecule, the screening by the plate makes retardation more rather
than less relevant to the force at small $z$, leading to an $\approx
10\%$ decrease in the force magnitude.

\begin{figure}[t!]
\centering
\includegraphics[width=0.9\columnwidth]{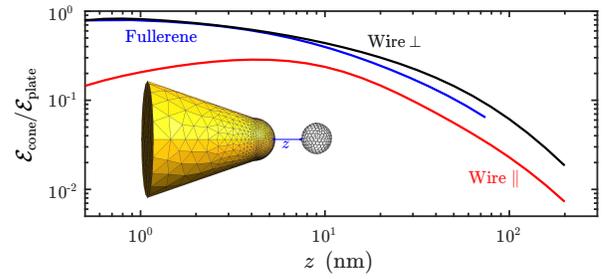}
\caption{Energy $\mathcal{E}_{\mathrm{cone}}$ of either a fullerene
  (solid blue) or carbyne wire (solid red/black) above a gold cone,
  normalized to the energy $\mathcal{E}_{\mathrm{plate}}$ of the same
  molecule but separated from a gold plate by the same
  surface--surface vertical distance $z$.}
\label{fig:cone}
\end{figure}

Finally, we consider the energy of a molecule above a gold conical
tip [\figref{cone}] by comparing it to that of a gold plate at the same
vertical separation $z$, with $\mathbb{G}_{\mathrm{env}}$ in the
former computed through the use of a free, surface-integral Maxwell
solver, SCUFF-EM~\cite{SCUFF1, SCUFF2}. The \emph{finite} cone has a
base diameter of $54~\mathrm{nm}$ and a height of $50~\mathrm{nm}$
from the base to the bottom of a hemispherical tip of diameter
$20~\mathrm{nm}$.  The ratio decreases with increasing $z$, with the
energy scaling as $z^{-6}$ at asymptotically large separations (not
shown) as the finite sizes of the cone and molecule become irrelevant
and their interactions dipolar. (Note that a decreasing ratio is
expected also for a semi-infinite cone due to its smaller effective
area and hence stronger decay compared to a plate.) The ratios at
small $z$ for the fullerene and perpendicular wire approach 1 since in
this limit, their small horizontal sizes allow the hemispherical tip,
which effectively acts like a plate at such short separations, to
dominate the interaction. By contrast, the ratio in the case of a
parallel wire is non-monotonic, decreasing with at short separations
since in this configuration, the wire excitations in the limit $z \to
0$ still sample the finite curvature of the tip and conical slope,
leading to a different asymptotic power law.

In conclusion, we have demonstrated a unifying approach to computing
vdW interactions among molecules and macroscopic bodies that accounts
for many-body and multiple-scattering effects to all orders. By
comparing against commonplace pairwise, CP, and non-retarded
approximations, we quantified the impact of nonlocality, finite size,
and retardation on the vdW energy between molecules and either a
planar or conical macroscopic body. We have consistently found larger
deviations in approximate interactions for long, semi-metallic
molecules such as carbyne wires, whereas compact, insulating molecules
such as many proteins are reasonably well-described as effectively
dilute dielectric particles, allowing these low-order approximations
to be more valid. In the future, one might consider more complex
macroscopic bodies, such as periodic gratings~\cite{BenderPRX2014,
  BuhmannIJMPA2016} that may elicit larger differences between RMB and
approximate interactions even for compact biomolecules, as well as
extend these results to incorporate the effects of infrared molecular
resonances~\cite{BuhmannPRA2012}.

This material is based upon work supported by the National Science
Foundation under Grant No. DMR-1454836 and by the National Science
Foundation Graduate Research Fellowship Program under Grant No. DGE
1148900.

%Our general formalism would allow us to further investigate
%interactions among molecules above more complicated continuous bodies,
%such as periodic gratings~\cite{BenderPRX2014,
%  BuhmannIJMPA2016}. Farther in the future, we aim to characterize vdW
%interactions that include the effects of infrared (surface phonon)
%resonances in molecules, incorporating changes in the susceptibility
%within our framework in a manner akin previous
%work~\cite{BuhmannPRA2012}. Additionally, we hope to be able to
%incorporate long-range dispersive electromagnetic effects in the
%presence of macroscopic dielectric bodies into the ab-initio DFT
%computation of atomic polarizabilities, as currently those
%polarizabilities are computed in vacuum without
%retardation~\cite{DobsonJPCM2012, DiStasioJPCM2014}.

\appendix

\section{Approximations}

Our calculations above make two related approximations related to
Gaussian damping. First, we approximate~\eqref{Gelem} as
\begin{equation}
  \langle f_{p}|\mathbb{G} \mathbb{V}_{k} f_{q} \rangle
  \approx \alpha_{q} \int~\mathrm{d}^{3} \vec{x}'~\mathbb{G} (\im\xi,
  \vec{x}_{p}, \vec{x}') f_{p+q} (\im\xi, \vec{x}')
\end{equation}                                                               
where $f_{p + q}$ is the same as $f_{q}$, but with $\sqrt{2}
\sigma_{p}$ replaced with $\sqrt{\sigma_{p}^{2} + \sigma_{q}^{2}}$, in
line with~\cite{DiStasioJPCM2014}; this effectively approximates the
Galerkin discretization by a collocation method, with the basis
Gaussian functions $f_{p + q}$ acquiring modified widths. Secondly,
for computational convenience, we consider only scattered fields
(Green's functions) from dipolar rather than Gaussian sources, which
is justified so long as the atoms are several widths (angstroms) away
from the macroscopic surfaces.

\section{vdW energy via fluctuation--dissipation theorem}

We provide a heuristic derivation of the retarded, many-body (RMB) vdW
energy of a general collection of molecular or macroscopic bodies,
requiring only that they be disjoint and have no correlations in the
polarization response \emph{between} bodies. Each body $k$ is
described by an electric susceptibility% (``scattering potential'')
$\mathbb{V}_{k}$, relating its polarization to the total electric
field via $\ket{\vec{P}_{k}} = \mathbb{V}_{k} \ket{\vec{E}}$; these
susceptibilities account for short-range quantum and electrostatic
correlations, allowing us to focus solely on long-range electrodynamic
correlations when considering the vdW energy.~\footnote{When
  neglecting retardation, the \emph{charge} density and electric
  potential are more frequently used, so the density response is
  written in terms of the susceptibility as $\chi(\omega, \vec{x},
  \vec{x}') = \sum_{i,j} \partial_{i} \partial_{j}
  (\mathbb{V}_{k})_{ij} (\omega, \vec{x}, \vec{x}')$.} Our derivation
follows analysis~\cite{RosaPRA2011, AgarwalPRA1975, MacRuryJCP1973}
based on the fluctuation--dissipation theorem; we note previous
demonstrations~\cite{MahantyJPAGP1972, TkatchenkoJCP2013} of its
equivalence to the summation of ground-state energies of the coupled
molecular system~\cite{RennePHYSICA1971A, RennePHYSICA1971B}.

Following~\citeasnoun{RosaPRA2011}, the assembly of the constituents
of all bodies from infinite separation into the final configuration
defining $\mathbb{V} = \sum_{k} \mathbb{V}_{k}$ can be considered the
result of an adiabatic change in the particle--field coupling strength
$\lambda \in [0, 1]$, in which case the energy of the system can be
written as,
\begin{equation}
  \mathcal{E} = -\int_{0}^{\infty}~\mathrm{d}\omega~\int_{0}^{1}~\frac{\mathrm{d}
    \lambda}{\lambda}~\bracket{\bracket{\vec{P}|\vec{E}}}_{\mathrm{ZP}},
\end{equation}
per the Feynman--Hellmann theorem~\cite{RosaPRA2011,
  AgarwalPRA1975}. Here, $\ket{\vec{E}}$ denotes zero-point
fluctuating electric fields, $\ket{\vec{P}} = \mathbb{V}
\ket{\vec{E}}$ is the induced polarization, and
$\bracket{~}_{\mathrm{ZP}}$ denotes the quantum statistical average
over zero-point fluctuations. The connection to scattering problems
comes from the well-known fluctuation--dissipation
theorem~\cite{Johnson2011},
\begin{equation}
  \bracket{\ket{\vec{E}} \bra{\vec{E}}}_{\mathrm{ZP}} = 
  \frac{\hbar}{\pi} \Im \mathbb{G},
%\left(\mathbb{I} - \mathbb{G}_{0} \mathbb{V}\right)^{-1} \mathbb{G}_{0}
\end{equation}
which expresses field fluctuations in terms of the Green's function
$\mathbb{G}$ of the system. The latter solves Maxwell's equations and
can be written in terms of the susceptibility as $\mathbb{G} =
(\mathbb{I} - \mathbb{G}_{0} \mathbb{V})^{-1}
\mathbb{G}_{0}$~\cite{RahiPRD2009}. Exploiting the analyticity of
$\mathbb{V}$ and $\mathbb{G}_{0}$ in the complex-$\omega$
plane~\cite{RodriguezNATURE2011, RodriguezPRA2007} and performing a
Wick rotation of the energy integral from real to imaginary frequency
$\omega = \im\xi$, leads to a simplified expression for the
energy~\footnote{The entirety of this derivation is identical to that
  of past work employing the so-called adiabatic connection
  fluctuation--dissipation (ACFD) framework, but using the vector
  polarization, tensor electric susceptibility, and tensorial vacuum
  Green's function instead of the scalar charge density, density
  response, or Coulomb potential, in order to account for
  retardation. It is therefore not a coincidence that the
  log-determinant frequency integrand is so similar in form to past
  expressions for the vdW energy of a single body.},
\begin{equation} \label{eq:Uvac}
  \mathcal{E} = \frac{\hbar}{2\pi}
  \int_{0}^{\infty}~\mathrm{d}\xi~\ln\left(\det\left(\mathbb{I} - \mathbb{G}_{0}
      \mathbb{V}\right)\right)
\end{equation}
where we rescaled the response functions $\mathbb{G}_{0}$ and
$\mathbb{V}$ by the coupling constant $\lambda$ and integrated over
$\lambda$. The net interaction energy among the bodies is found by
subtracting self-energies of the form in~\eqref{Uvac}, replacing
$\mathbb{V}$ by $\mathbb{V}_{k}$ separately for each $k$. This allows
recasting the net vdW interaction energy as~\eqref{MM} in terms of
scattering operators:
\begin{align}
  \mathbb{M} &= \mathbb{I} - \mathbb{G}_{0} \mathbb{V} \\
  \mathbb{M}_{\infty} &= \prod_{k} (\mathbb{I} - \mathbb{G}_{0}
  \mathbb{V}_{k}).
\end{align}

If the system considered consists of $N$ molecular bodies labeled $k$,
and an arbitrary number of macroscopic bodies collectively described
by $\mathbb{V}_{\mathrm{env}}$, then one can write
\begin{align}
  \mathbb{M} &= \mathbb{I} - \mathbb{G}_{0} \mathbb{V}_{\mathrm{env}}
  - \mathbb{G}_{0} \mathbb{V} \\
  \mathbb{M}_{\infty} &= (\mathbb{I} - \mathbb{G}_{0}
  \mathbb{V}_{\mathrm{env}}) \prod_{k} (\mathbb{I} - \mathbb{G}_{0}
  \mathbb{V}_{k}),
\end{align}
where $\mathbb{V} = \sum_{k} \mathbb{V}_{k}$ only runs over the
molecular bodies. Multiplying $\mathbb{M}\mathbb{M}_{\infty}^{-1}$
produces terms of the form
\begin{equation}
  (\mathbb{I} - \mathbb{G}_{0}
  \mathbb{V}_{\mathrm{env}})^{-1} \mathbb{G}_{0}  \equiv \mathbb{G}_{\mathrm{env}},
\end{equation}
which is just the electric field response due to
$\mathbb{V}_{\mathrm{env}}$ alone and can be computed via analytical
or numerical formulations of continuum electrodynamics. Redefining
\begin{align}
  \mathbb{M}' &= \mathbb{I} - \mathbb{G}_{\mathrm{env}} \mathbb{V} \\
  \mathbb{M}'_{\infty} &= \prod_{k} (\mathbb{I} - \mathbb{G}_{0}
  \mathbb{V}_{k}),
\end{align}
and dropping primes, these new operators can then be substituted
into~\eqref{MM} to obtain the net vdW interaction energy among $N$
molecules and a general macroscopic environment.

\bibliography{molvdwplanepaper}
\end{document}